\documentclass[aps,pre,twocolumn,superscriptaddress]{revtex4-1}
\usepackage{graphicx}
\usepackage{dcolumn}
\usepackage{bm}
\usepackage{amsmath}
\usepackage{amssymb}

\begin{document}

\title{Estimation of anisotropic bending rigidities and spontaneous curvatures of crescent curvature-inducing proteins from tethered-vesicle experimental data}

\author{Hiroshi Noguchi}
\email[]{noguchi@issp.u-tokyo.ac.jp}
\affiliation{Institute for Solid State Physics, University of Tokyo, Kashiwa, Chiba 277-8581, Japan}
\author{Nikhil Walani}
\affiliation{Universitat Polit{\`e}dcnica de Catalunya-BarcelonaTech,
08034 Barcelona, Spain}
\affiliation{Present address: Department of Applied Mechanics, IIT Delhi, Hauz Khas, New Delhi, 110016, India}
\author{Marino Arroyo}
\affiliation{Universitat Polit{\`e}dcnica de Catalunya-BarcelonaTech,
08034 Barcelona, Spain}
\affiliation{Institute for Bioengineering of Catalonia (IBEC), The Barcelona Institute for
Science and Technology (BIST), 08028 Barcelona, Spain}
\affiliation{Centre Internacional de M{\`e}todes Num{\`e}rics en Enginyeria (CIMNE), 08034
Barcelona, Spain}

\begin{abstract}
The Bin/amphiphysin/Rvs (BAR) superfamily proteins
have a crescent binding domain and bend biomembranes along the domain axis.
However, their anisotropic bending rigidities and spontaneous curvatures have not been experimentally determined.
Here, we estimated these values from the bound protein densities on tethered vesicles 
using a mean-field theory of anisotropic bending energy 
and orientation-dependent excluded volume.
The dependence curves of the protein density on the membrane curvature are fitted to the experimental data for the I-BAR and N-BAR domains 
reported by  C. Pr{\'e}vost et al. Nat. Commun. {\bf 6}, 8529 (2015) and F.-C. Tsai et al. Soft Matter {\bf 17}, 4254 (2021), respectively.
For the I-BAR domain, all three density curves of different chemical potentials exhibit excellent fits
with a single parameter set of anisotropic bending energy.
When the classical isotropic bending energy is used instead, one of the curves can be fitted well, but the others exhibit large deviations.
In contrast, for the N-BAR domain, two curves are not well-fitted simultaneously using the anisotropic model, 
although it is significantly improved compared to the isotropic model.
This deviation likely suggests a cluster formation of the N-BAR domains.
\end{abstract}

\maketitle

\section{Introduction}

In living cells, membrane morphology is regulated by the binding and unbinding of
 curvature-inducing proteins~\cite{mcma05,suet14,joha15,bran13,hurl10,mcma11,baum11,has21}.
Some types of these proteins bend a membrane in a laterally isotropic manner
and generate  spherical membrane buds~\cite{joha15,bran13,hurl10,mcma11}.
In contrast,  the Bin/amphiphysin/Rvs (BAR) superfamily proteins have a crescent binding domain (BAR domain)
and bend membranes along the BAR domain axis,
generating cylindrical membrane tubes~\cite{mcma05,suet14,joha15,itoh06,masu10,mim12a,fros08}.
Several types of BAR domains are known:
N-BARs and F-BARs bend membranes positively, but I-BARs bend them in the opposite direction.

These curvature-inducing proteins can sense membrane curvature; that is,
their binding onto membranes depends on the local membrane curvatures.
Tethered vesicles have been widely used to observe the curvature sensing experimentally~\cite{baum11,has21,sorr12,prev15,tsai21,rosh17,aimo14,yang22,roux10,more19,lars20}.
A vesicle is pulled by optical tweezers and a micropipette
to form a narrow membrane tube (tether).
The tube radius can be controlled by adjusting the position of the optical tweezers.
The curvature sensing of
BAR proteins~\cite{baum11,has21,sorr12,prev15,tsai21}, G-protein coupled receptors (GPCRs)~\cite{rosh17}, ion channels~\cite{aimo14,yang22}, dynamin~\cite{roux10},  annexins~\cite{more19}, and Ras proteins~\cite{lars20}
have been reported.
Additionally, the curvature sensing has been detected by the protein binding onto different sizes of spherical vesicles~\cite{lars20,hatz09,zeno19}.

Evaluating the mechanical properties of these proteins is crucial for quantitatively understanding their curvature generation and sensing.
The aim of this study is to determine the anisotropic bending rigidity and spontaneous curvature of BAR proteins from experimental data of tethered vesicles.
In previous studies~\cite{sorr12,prev15,tsai21}, the bending rigidity and spontaneous curvature of BAR proteins have been estimated using the Canham--Helfrich theory~\cite{canh70,helf73}.
However, this theory is formulated for laterally isotropic fluid membranes; thus,
the anisotropy of the proteins is not considered.
Recently, we developed a mean-field model for anisotropic bending energy and entropic interactions~\cite{tozz21,roux21,nogu22}.
Orientational fluctuations are included based on Nascimentos' theory for three-dimensional liquid crystals~\cite{nasc17}.
The first- and second-order transitions between isotropic and nematic phases are obtained with increasing protein density in narrow membrane tubes~\cite{nogu22}.
In the present study, we use this theoretical model to estimate the anisotropic bending rigidity and spontaneous curvature.
The experimental data for the I-BAR domain of IRSp53 and the N-BAR domain of amphiphysin 1 reported in Refs.~\citenum{prev15} and \citenum{tsai21}, respectively,
are used for the estimation.
In theoretical studies, a protein is often assumed to be a rigid body~\cite{domm99,domm02,schw15,kohy19a}.
The membrane-mediated interaction of two rigid proteins qualitatively reproduces that of two flexible proteins obtained by meshless membrane simulations,
but the amplitude is overestimated~\cite{nogu17}.
Hence, the estimation of the bending rigidity is also important for evaluating the interaction between proteins.

The mean-field theory and fitting method are described in Sec.~\ref{sec:theory}.
Secs.~\ref{sec:ibar0} and \ref{sec:nbar0} present and discuss
the fitting results for the I-BAR and N-BAR domains, respectively.
Additionally, the results of the isotropic and anisotropic protein models are compared.
Sec.~\ref{sec:sum} concludes the paper.

\begin{figure}[tbh]
\includegraphics[]{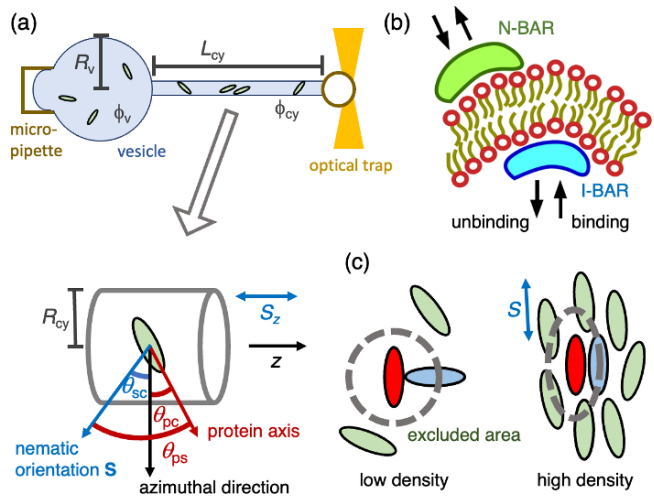}
\caption{
Schematic of a tethered vesicle and protein binding.
(a) Experimental setup of the tethered vesicle.
The proteins are bound in the tether and spherical vesicle regions
with bound densities of $\phi_{\mathrm {cy}}$ and $\phi_{\mathrm {v}}$, respectively.
The angles between the nematic direction {\bf S}, the azimuthal direction, and/or
the protein axis are shown in the bottom panel.
(b) Binding and unbinding of BAR domains.
N-BAR and I-BAR domains bind onto the outer and inner surfaces of the vesicle, respectively.
(c) Excluded-volume interactions between proteins.
A perpendicular protein pair has a larger excluded area on the membrane surface (represented by thick dashed lines)
than a parallel pair (compare the left and right panels).
}
\label{fig:cat}
\end{figure}

\section{Theory}\label{sec:theory}

A cylindrical membrane tube (tether) protrudes from
the spherical vesicle, as depicted in Fig.~\ref{fig:cat}(a).
The tether length $L_{\mathrm {cy}}$ and radius $R_{\mathrm {cy}}$ are controlled by force $f_{\mathrm{ex}}$ 
generated by optical tweezers and a micropipette.
The membrane is in a fluid phase and is homogeneous.
The radius $R_{\mathrm {v}}$ of the spherical region is on a $\mu$m scale; thus,
the membrane can be approximated as flat.
The subscripts v and cy represent the quantities in the spherical and tether regions, respectively.
The total membrane area $A$ is fixed and the membrane area inside of the micropipette is assumed to be constant.
The tether area $A_{\mathrm {cy}}= 2\pi R_{\mathrm {cy}}L_{\mathrm {cy}}$ is approximated as a constant,
since the tube volume is negligibly small~\cite{smit04,nogu21b}.
The protein density  $\phi$ is the local area fraction covered by the bound proteins
($\phi_{\mathrm {v}}$ and $\phi_{\mathrm {cy}}$ represent the densities in the spherical and tether regions, respectively).
N-BAR and I-BAR proteins bind onto the outer and inner surfaces, respectively (see Fig.~\ref{fig:cat}(b)).
Here, the curvature direction is defined as outward following the membrane curvature.
Hence, the proteins binding to the inner surface have the opposite sign of curvature 
from the protein viewpoint (see I-BAR in Fig.~\ref{fig:cat}(b)).

\subsection{Isotropic proteins}\label{sec:iso}

First, we describe the mean-field theory of proteins that bend membranes isotropically (no preferred lateral direction).
The bending energy is given as follows:~\cite{nogu21a,nogu22a}
\begin{eqnarray}\label{eq:Fcv0}
F_{\mathrm {cv}} &=&  4\pi\bar{\kappa}_{\mathrm {d}}(1-g_{\mathrm {ves}}) + \int {\mathrm {d}}A \ \Big\{ 2\kappa_{\mathrm {d}}H^2(1-\phi) \nonumber \\
&&  + \frac{\kappa_{\mathrm {pi}}}{2}(2H-C_0)^2\phi  
+ (\bar{\kappa}_{\mathrm {pi}}-\bar{\kappa}_{\mathrm {d}})K \phi  \Big\},
\end{eqnarray}
where $g_{\mathrm {ves}}$ represents the genus of the vesicle ($g_{\mathrm {ves}}=0$ for tethered vesicles).
 $H=(C_1+C_2)/2$ and $K=C_1C_2$ represent the mean and Gaussian curvatures of each position, respectively,
with $C_1$ and $C_2$ being the principal curvatures.
The bare (protein-unbound) membrane has a bending rigidity of $\kappa_{\mathrm {d}}$, zero spontaneous curvature, 
and saddle-splay modulus of $\bar{\kappa}_{\mathrm {d}}$ (also called the Gaussian modulus) in the Canham--Helfrich theory~\cite{canh70,helf73,safr94}.
The bound membrane has a bending rigidity of $\kappa_{\mathrm {pi}}$, finite spontaneous curvature $C_0$, and saddle-splay modulus of $\bar{\kappa}_{\mathrm {pi}}$.
The first term of eqn~(\ref{eq:Fcv0}) represents the integral over the Gaussian curvature $K$.
Note that the curvature mismatch model~\cite{tsai21,has21,prev15,rosh17} and spontaneous curvature model~\cite{tsai21,has21,rama00,tozz19}
are subsets of the present model for $\kappa_{\mathrm {pi}}>\kappa_{\mathrm {d}}$ and $\kappa_{\mathrm {pi}}=\kappa_{\mathrm {d}}$, respectively~\cite{nogu21a}.
For $\kappa_{\mathrm {pi}}<\kappa_{\mathrm {d}}$, the proteins exhibit curvature sensing but do not have a non-passive curvature-generation capability~\cite{nogu22a}.

The membrane free energy $F$ consists of 
the binding energy and mixing entropy in addition to the bending energy $F_{\rm cv}$,
\begin{equation}\label{eq:F0}
F =  F_{\mathrm {cv}} + \int {\mathrm {d}}A \ \Big\{ - \frac{\mu}{a_{\mathrm {p}}}\phi +
  \frac{k_{\mathrm {B}}T}{a_{\mathrm {p}}}[\phi \ln(\phi) +  (1-\phi) \ln(1-\phi) ] \Big\},
\end{equation}
where 
 $a_{\mathrm {p}}$ represents the area covered by one protein and  $k_{\mathrm {B}}T$ represents the thermal energy.
The maximum number of bound proteins is $A/a_{\mathrm {p}}$.
The first and second terms in the integral of eqn~(\ref{eq:F0}) represent the protein-binding energy
with the chemical potential $\mu$ and the mixing entropy of bound proteins, respectively.
Here, we neglect the inter-protein interaction energy ($\sim \phi^2$)~\cite{nogu21a,nogu21b,nogu22a},
since we consider low protein densities in this study.

In thermal equilibrium,
the protein density $\phi$ is locally determined for each membrane curvature:~\cite{nogu21a,nogu22a}
\begin{eqnarray}\label{eq:phi0}
\phi &=& \frac{1}{1+\exp(w_{\mathrm {b}})}, \\
w_{\mathrm {b}}&=& - \frac{\mu}{k_{\mathrm {B}}T} \\ \nonumber && + \frac{a_{\mathrm {p}}}{k_{\mathrm {B}}T}\Big(2\kappa_{\mathrm {dif}}H^2 + \bar{\kappa}_{\mathrm {dif}}K 
 -2\kappa_{\mathrm {pi}}C_0H  + \frac{\kappa_{\mathrm {pi}}C_0^2}{2} \Big),
\end{eqnarray}
where $\kappa_{\mathrm {dif}}=\kappa_{\mathrm {pi}}-\kappa_{\mathrm {d}}$
and $\bar{\kappa}_{\mathrm {dif}}=\bar{\kappa}_{\mathrm {pi}}-\bar{\kappa}_{\mathrm {d}}$.
Since the curvature of the spherical region of the tethered vesicles is approximated as $H=K=0$, 
the protein density in the spherical region is given as
$\phi_{\mathrm{v}} = 1/\{1+\exp[(-\mu + a_{\mathrm {p}}\kappa_{\mathrm {pi}}C_0^2/2)/k_{\mathrm {B}}T]\}$.
Hence, the protein density $\phi_{\mathrm{cy}}$ in the tether regions
is given as 
\begin{equation}\label{eq:phiia}
\phi_{\mathrm{cy}} = \frac{1}{1+ \frac{1-\phi_{\mathrm{v}}}{\phi_{\mathrm{v}}}\exp\big[\frac{a_{\mathrm {p}}}{k_{\mathrm {B}}T}\big(\frac{\kappa_{\mathrm {dif}}}{{2R_{\mathrm{cy}}}^2}
 -\frac{\kappa_{\mathrm {pi}}C_0}{R_{\mathrm{cy}}} \big)\big] }.
\end{equation}
When the membrane has
the sensing curvature $C_{\mathrm{s}}$,
$\phi_{\mathrm{cy}}$ is maximized. 
The sensing curvature is obtained from $\phi_{\mathrm{cy}}/(1/R_{\mathrm{cy}})=0$:
\begin{equation}\label{eq:cs}
C_{\mathrm{s}}= \frac{\kappa_{\mathrm {pi}}C_0}{\kappa_{\mathrm {dif}}}.
\end{equation}
For $\kappa_{\mathrm {dif}} \ne 0$, the numerator of the second term in parentheses 
can be replaced with $\kappa_{\mathrm {dif}}C_{\mathrm{s}}$,
so that $\kappa_{\mathrm {dif}}$ and $C_{\mathrm{s}}$ can be used as fitting parameters.
Here, the protein densities of tethered vesicles are independent of $\bar{\kappa}_{\mathrm {dif}}$,
since $R_{\mathrm {v}} \gg R_{\mathrm {cy}}$.
Note that at $R_{\mathrm {v}} \sim R_{\mathrm {cy}}$,
$\phi_{\mathrm{v}}$ is also dependent on $\bar{\kappa}_{\mathrm {dif}}$~\cite{nogu22a}.
Thus, the density difference between small vesicles and tethers of the same mean curvature reported in Ref.~\citenum{lars20}
may be caused by this Gaussian curvature dependence.
Other isotropic proteins, such as GPCRs~\cite{rosh17} and ion channels~\cite{aimo14,yang22}, may also have similar dependences.

For the low-density limit ($\phi_{\mathrm{cy}}\ll 1$),
the density ratio is expressed by the exponential function~\cite{prev15,nogu21b}
\begin{equation}\label{eq:phiis}
\frac{\phi_{\mathrm{cy}}}{\phi_{\mathrm{v}}} = \exp\Big[ -\frac{a_{\mathrm {p}}}{k_{\mathrm {B}}T}\Big(\frac{\kappa_{\mathrm {dif}}}{{2R_{\mathrm{cy}}}^2}
 -\frac{\kappa_{\mathrm {pi}}C_0}{R_{\mathrm{cy}}} \Big)\Big].
\end{equation}
The ratio $\phi_{\mathrm{cy}}/\phi_{\mathrm{v}}$ is independent of $\phi_{\mathrm{v}}$ in this limit.
Details regarding isotropic-protein binding on tethered vesicles are described in Ref.~\citenum{nogu21b}.

\subsection{Anisotropic proteins}\label{sec:aniso}

Anisotropies of the protein bending energy and excluded volume are considered.
The lateral shape of a bound protein is approximated as an ellipse
with  major and minor axis lengths of $\ell_1$ and $\ell_2$, respectively.
The aspect ratio is $d_{\mathrm {el}}=\ell_1/\ell_2$, and the area is $a_{\mathrm {p}} = \pi \ell_1\ell_2/4$.
These proteins have an orientation-dependent excluded-volume interaction and
can align on the membrane surface.
When neighboring proteins have a perpendicularly orientation,
the excluded area $A_{\mathrm {exc}}$ between them is larger than that for parallel pairs, as shown in Fig.~\ref{fig:cat}(c).
This area  $A_{\mathrm {exc}}$ is approximated as a function of the angle $\theta_{\mathrm {pp}}$ between the major axes of the two proteins:~\cite{nogu22}
 $A_{\mathrm {exc}}= a_{\mathrm {p}}[4 - b_{\mathrm {exc}}(\cos^2(\theta_{\mathrm {pp}})-1)]$.
The effective excluded area is $A_{\mathrm {eff}}= \lambda A_{\mathrm {exc}}$.
Although $\lambda$ decreases  slightly with an increase in the protein density,
we use a constant value $\lambda = 1/3$ for simplicity~\cite{tozz21,roux21,nogu22}.

The bending energy of a bound protein is given as follows:
\begin{eqnarray}
U_{\mathrm {p}} &=&   \frac{\kappa_{\mathrm {p}}a_{\mathrm {p}}}{2}(C_{\ell 1} - C_{\mathrm {p}})^2 + \frac{\kappa_{\mathrm {side}}a_{\mathrm {p}}}{2}(C_{\ell 2} - C_{\mathrm {side}})^2, \\
C_{\ell 1} &=& C_1 \cos^2( \theta_{\mathrm {pc}} ) + C_2  \sin^2( \theta_{\mathrm {pc}}), \\
C_{\ell 2} &=& C_1 \sin^2( \theta_{\mathrm {pc}} ) + C_2  \cos^2(\theta_{\mathrm {pc}}),
\end{eqnarray}
where $C_{\ell 1}$ and $C_{\ell 2}$ represent the curvatures along the major and minor axes of the protein, respectively,
and $\theta_{\mathrm {pc}}$ represents the angle between the major protein axis and membrane principal direction
(the azimuthal direction of the cylindrical tube), as shown in Fig.~\ref{fig:cat}(a).
The proteins can have  different values of bending rigidity and spontaneous curvature along the major and minor protein axes:
$\kappa_{\mathrm {p}}$ and $C_{\mathrm {p}}$ along the major protein axis
and  $\kappa_{\mathrm {side}}$ and $C_{\mathrm {side}}$ are along the minor axis (side direction).

The free energy $F_{\mathrm {p}}$ of the bound proteins is expressed as follows:
\begin{eqnarray}
F_{\mathrm {p}} &=& \int f_{\mathrm {p}}\ {\mathrm{d}}A, \\ 
f_{\mathrm {p}} &=&  \frac{\phi k_{\mathrm {B}}T}{a_{\mathrm {p}}}\Big[\ln(\phi) + \frac{S \Psi}{2} - \ln\Big(\int_{-\pi}^{\pi} w(\theta_{\mathrm {ps}})\ {\mathrm{d}}\theta_{\mathrm {ps}}\Big)\Big],\hspace{0.5cm} \\
w(\theta_{\mathrm {ps}})  &=&  g\exp\Big[\Psi s_{\mathrm {p}}(\theta_{\mathrm {ps}}) + \bar{\Psi}\sin(\theta_{\mathrm {ps}})\cos(\theta_{\mathrm {ps}}) \nonumber \\
 && - \frac{U_{\mathrm {p}}}{k_{\mathrm {B}}T} \Big]\Theta(g), \\
g   &=& 1-\phi (b_0-b_2S s_{\mathrm {p}}(\theta_{\rm ps})),
\end{eqnarray}
where 
$\Theta(x)$ denotes the unit step function and $s_{\mathrm {p}}(\theta_{\mathrm {ps}}) = \cos^2(\theta_{\mathrm {ps}}) - 1/2$.
The proteins are ordered as $S = 2 \langle s_{\mathrm {p}}(\theta_{\mathrm {ps}}) \rangle$,
where $\theta_{\mathrm {ps}}$ represents the angle between the major protein axis and  the ordered direction
and $\langle ... \rangle$ denotes the ensemble average (see Fig.~\ref{fig:cat}).
Factor $g$ expresses the effect of the orientation-dependent excluded volume, where 
 $b_0= (4 + b_{\mathrm {exc}}/2)\lambda$ and $b_2= b_{\mathrm {exc}}\lambda$.
At $d_{\rm el}=2$, $3$, $4$, and $6$, $b_{\mathrm {exc}}= 0.840$, $1.98$, $3.44$, and $6.14$, respectively.
Non-overlapped states exist at $g>0$.
The ensemble average of a protein quantity $\chi$ is given as
\begin{eqnarray} \label{eq:av}
\langle \chi \rangle = \frac{\int_{-\pi}^{\pi} \chi w(\theta_{\mathrm {ps}})\ {\mathrm{d}}\theta_{\mathrm {ps}} }{\int_{-\pi}^{\pi}  w(\theta_{\mathrm {ps}}) \ {\mathrm{d}}\theta_{\mathrm {ps}}}.
\end{eqnarray}
The quantities $\Psi$ and $\bar{\Psi}$ are the symmetric  and asymmetric components of the nematic tensor, respectively, 
and are determined using $S$ and $\langle \sin(\theta_{\mathrm {ps}})\cos(\theta_{\mathrm {ps}}) \rangle =0$ via eqn~(\ref{eq:av}).
When the nematic order is parallel to one of the directions of the membrane principal curvatures ($\theta_{\mathrm {sc}}=0$ or $\pi/2$), $\bar{\Psi}=0$.
In this study, the integral is performed in the range of $-\pi< \theta_{\mathrm {ps}} \le\pi$.
Since the shape is rotationally symmetric, the range $-\pi/2< \theta_{\mathrm {ps}} \le\pi/2$ can be used alternatively,
 in which the chemical potential is shifted by $\Delta\mu=k_{\mathrm{B}}T\ln(2)$~\cite{footnote1}.
Note that the separate integrals for the bending energy and other terms used in Ref.~\citenum{vyas22} are not applicable,
since the orientational fluctuations of proteins are significantly large~\cite{tozz21,nogu22}.

Since an external force $f_{\mathrm {ex}}$ is imposed, 
the free energy of the membrane tether is given as $F=F_{\mathrm {p}} + U_{\mathrm {mb}} - f_{\mathrm {ex}}L_{\mathrm {cy}}$,
where the energy of the bare (unbound) membrane is $U_{\mathrm {mb}}= \kappa_{\mathrm {d}}A/2{R_{\mathrm {cy}}}^2$.
This force $f_{\mathrm {ex}}$ is balanced with the membrane axial force and
 is obtained using $\partial F/\partial L_{\mathrm {cy}}|_{\phi} =0$, as follows:
\begin{eqnarray}
\label{eq:fex}
f_{\mathrm {ex}} = 2\pi\frac{\partial f_{\mathrm {p}}}{\partial (1/R_{\mathrm {cy}})}\bigg|_{\phi} + f_{\mathrm {mb}}.
\end{eqnarray}
Here, the last term $f_{\mathrm {mb}}$ represents the force of the bare membrane tube:
$f_{\mathrm {mb}}= 2\pi\kappa_{\mathrm {d}}/R_{\mathrm {cy}}$.

The equilibrium of binding and unbinding is obtained by
minimizing $F - \mu N_{\mathrm {p}}$, where $\mu$ represents the binding chemical potential.
Thus, the protein density is balanced at $\mu = a_{\mathrm {p}}\partial f_{\mathrm {p}}/\partial \phi$.
Details are described in Refs.~\citenum{tozz21} and \citenum{nogu22}.

\begin{figure}[tbh]
\includegraphics[]{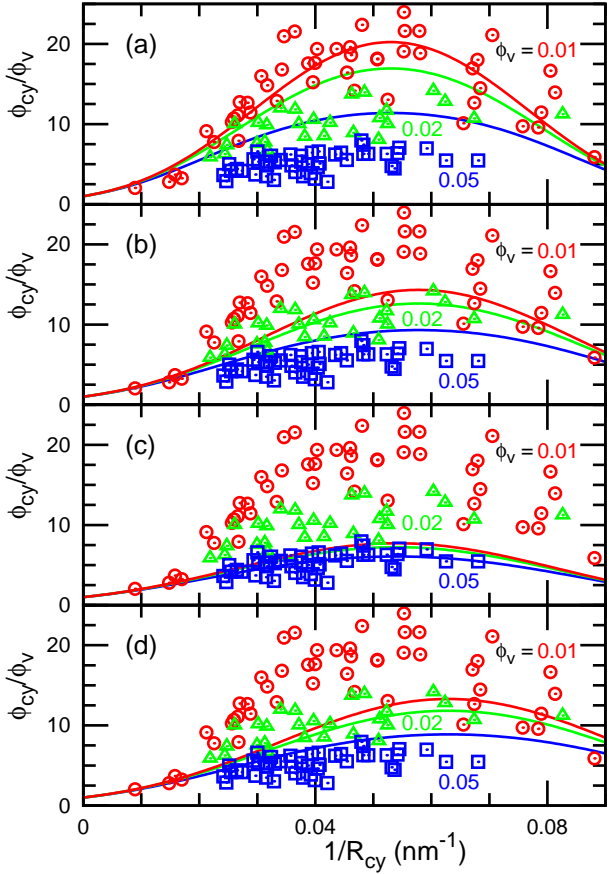}
\caption{
Fitting of the density--curvature curves for I-BAR-domain binding based on the isotropic protein model.
The density $\phi_{\mathrm{cy}}$ is normalized by $\phi_{\mathrm{v}}$ 
for the curvature $1/R_{\mathrm{cy}}$ of the tethered membrane.
Circles, triangles, and squares indicate
the experimental data for $\phi_{\mathrm{v}}=0.01$, $0.02$, and $0.05$, respectively (reproduced from Ref.~\citenum{prev15}).
The solid lines are given by eqn~(\ref{eq:phiia}) with the fitting parameters $\kappa_{\mathrm{dif}}$ and $C_{\mathrm{s}}$;
from top to bottom, $\phi_{\mathrm{v}}=0.01$, $0.02$, and $0.05$.
(a) Data at $\phi_{\mathrm{v}}=0.01$ are fitted:
 $\kappa_{\mathrm{dif}}/k_{\mathrm{B}}T=45.9$ and $C_{\mathrm{s}}=0.0530$\,nm$^{-1}$.
(b) Data at $\phi_{\mathrm{v}}=0.02$ are fitted:
 $\kappa_{\mathrm{dif}}/k_{\mathrm{B}}T=33.6$ and $C_{\mathrm{s}}=0.0578$\,nm$^{-1}$.
(c) Data at $\phi_{\mathrm{v}}=0.05$ are fitted:
 $\kappa_{\mathrm{dif}}/k_{\mathrm{B}}T=29.0$ and $C_{\mathrm{s}}=0.0540$\,nm$^{-1}$.
(d) All data are fitted:
 $\kappa_{\mathrm{dif}}/k_{\mathrm{B}}T=27.5$ and $C_{\mathrm{s}}=0.0629$\,nm$^{-1}$.
}
\label{fig:iiphi}
\end{figure}

\begin{figure}[tbh]
\includegraphics[]{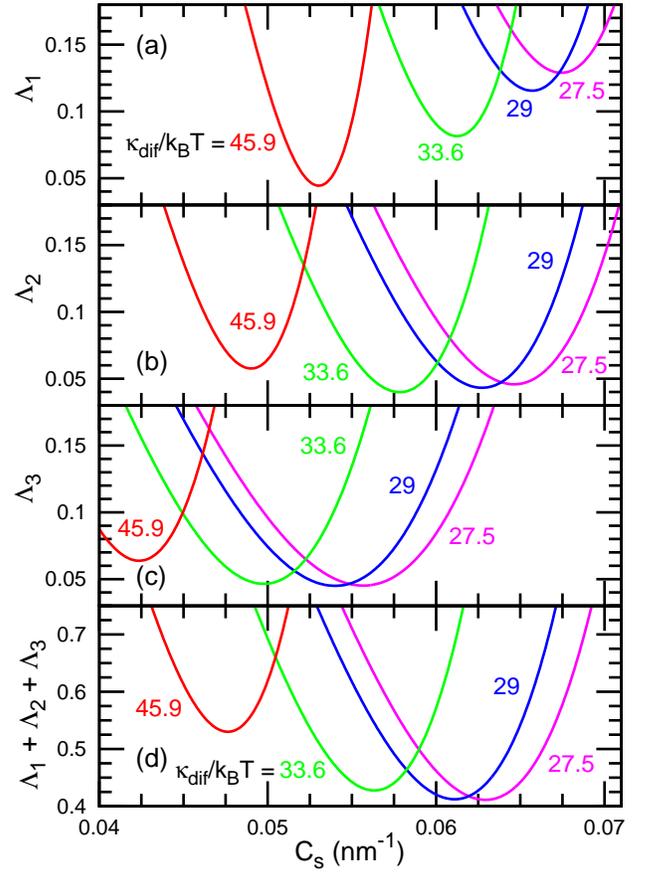}
\caption{
Fit deviation as a function of the sensing curvature $C_{\mathrm{s}}$ for I-BAR-domain binding based on the isotropic protein model.
From left to right,  $\kappa_{\mathrm{dif}}/k_{\mathrm{B}}T=45.9$, $33.6$, $29$, and $27.5$.
(a) Fit deviation $\Lambda_1$ for $\phi_{\mathrm{v}}=0.01$.
The minimum $\Lambda_1^{\mathrm{min}}=0.0444$ is obtained at $\kappa_{\mathrm{dif}}/k_{\mathrm{B}}T=45.9$ and $C_{\mathrm{s}}=0.0530$\,nm$^{-1}$ (corresponding to Fig.~\ref{fig:iiphi}(a)).
(b) Fit deviation $\Lambda_2$ for $\phi_{\mathrm{v}}=0.02$.
The minimum $\Lambda_2^{\mathrm{min}}=0.0399$ is obtained at $\kappa_{\mathrm{dif}}/k_{\mathrm{B}}T=33.6$ and $C_{\mathrm{s}}=0.0578$\,nm$^{-1}$ (corresponding to Fig.~\ref{fig:iiphi}(b)).
(c) Fit deviation $\Lambda_3$ for $\phi_{\mathrm{v}}=0.05$.
The minimum $\Lambda_3^{\mathrm{min}}=0.0450$ is obtained at $\kappa_{\mathrm{dif}}/k_{\mathrm{B}}T=29.0$ and $C_{\mathrm{s}}=0.0540$\,nm$^{-1}$ (corresponding to Fig.~\ref{fig:iiphi}(c)).
(d) Sum of fit deviations $\Lambda_1+\Lambda_2+\Lambda_3$.
The minimum $(\Lambda_1+\Lambda_2+\Lambda_3)^{\mathrm{min}}=0.411$ is obtained at $\kappa_{\mathrm{dif}}/k_{\mathrm{B}}T=27.5$ and $C_{\mathrm{s}}=0.0629$\,nm$^{-1}$ (corresponding to Fig.~\ref{fig:iiphi}(d)).
}
\label{fig:iiS}
\end{figure}

\begin{figure}[tbh]
\includegraphics[]{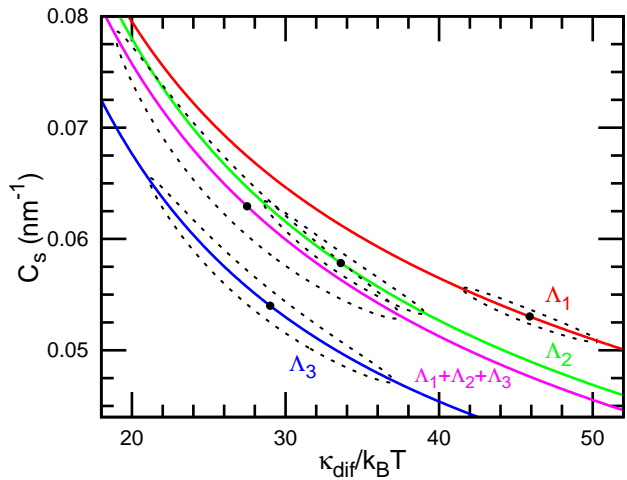}
\caption{
Two-dimensional map of the fit deviations $\Lambda_1$, $\Lambda_2$, $\Lambda_3$, and $\Lambda_1+\Lambda_2+\Lambda_3$ for I-BAR-domain binding based on the isotropic protein model.
The solid lines represent the valleys connecting the values of $C_{\mathrm{s}}$ for the lowest $\Lambda$ values at fixed $\kappa_{\mathrm{dif}}$.
The circles represent the minima and the dashed lines represent the contours of values exceeding the minima by $10$\%.
}
\label{fig:iimin}
\end{figure}

Since we consider a low density of bound proteins,
the proteins in the spherical vesicle region are randomly oriented, that is, $S=\Psi=\bar{\Psi}=0$.
Hence, the free energy density $f_{\mathrm {p,v}}$ of the spherical vesicle region is given as
\begin{equation}
f_{\mathrm {p,v}} =  \frac{\phi_{\mathrm {v}}}{a_{\mathrm {p}}}\Big\{ k_{\mathrm {B}}T\big[\ln(\phi_{\mathrm {v}}) - \ln(1-b_0\phi_{\mathrm {v}}) - \ln(2\pi)\big] + U_{\mathrm {p,v}}\Big\},
\end{equation}
where $U_{\mathrm {p,v}} = (\kappa_{\mathrm {p}}{C_{\mathrm {p}}}^2 + \kappa_{\mathrm {side}}{C_{\mathrm {side}}}^2)a_{\mathrm {p}}/2$.
Hence, $\mu$ for the density $\phi_{\mathrm {v}}$ is obtained as
\begin{equation}
\frac{\mu-U_{\mathrm {p,v}}}{k_{\mathrm {B}}T} =
\ln\bigg(\frac{\phi_{\mathrm {v}}}{1-b_0\phi_{\mathrm {v}}}\bigg) + \frac{b_0\phi_{\mathrm {v}}}{1-b_0\phi_{\mathrm {v}}} - \ln(2\pi) + 1.
\end{equation}

For the I-BAR and N-BAR domains, we use $d_{\mathrm {el}}=6$, and $3$, respectively.
For both proteins, we use $a_{\mathrm {p}}=50{\mathrm{\,nm}}^2$ in accordance with Refs.~\citenum{prev15} and \citenum{tsai21}.

We also calculate the orientational order $S_z$ along the tube ($z$) axis,
since it is more easily measured than $S$ in experiments.
When the orientational order is along the azimuthal and axial directions ($\theta_{\mathrm {sc}}=0$ and $\pi/2$), $S_z= - S$ and $S_z= S$, respectively.
At a high protein density and small tube radius ($1/R_{\mathrm{cy}}> C_{\mathrm{p}}$), 
the orientational order can deviate from the azimuthal or axial direction ($0<\theta_{\mathrm {sc}}<\pi/2$).
However, in this study, the fitted results remain in the range of $\theta_{\mathrm {sc}}=0$ and $\pi/2$, 
since the protein densities are sufficiently low.

\subsection{Fitting}\label{sec:fit}

The experimental data of the bound protein density on the membrane tether are used for the fitting.
We employ a least-squares method
and search the conditions for minimizing the mean squared deviation:
\begin{equation}
\Lambda = \frac{1}{{\phi_{\mathrm{m}}}^2N} \sum_i^N ( \phi_i - \phi_{\mathrm{theory}})^2, 
\end{equation}
where $N$ represents the number of experimental data, and
$\phi_i$ and $\phi_{\mathrm{theory}}$ represent the experimental and theoretical values of the protein density of the tether region, respectively.
This fit deviation is normalized by the mean value $\phi_{\mathrm{m}} = (1/N)\sum_i \phi_i$ of the experimental data.
If no normalization is applied, the obtained values of $\Lambda$ depend on the choice of units ($\phi_{\mathrm {cy}}$ or $\phi_{\mathrm {cy}}/\phi_{\mathrm {v}}$).
For the isotropic protein model, two fitting parameters, $\kappa_{\mathrm{dif}}$ and $C_{\mathrm{0}}$, are used.
For the anisotropic protein model,
two fitting parameters, $\kappa_{\mathrm{p}}$ and $C_{\mathrm{p}}$,
or four fitting parameters, $\kappa_{\mathrm{p}}$, $C_{\mathrm{p}}$, $\kappa_{\mathrm{side}}$, and $C_{\mathrm{side}}$, are used.

For the I-BAR domain of IRSp53,
the experimental data  reported in Ref.~\citenum{prev15} are used.
The fit deviations for $\phi_{\mathrm{v}}=0.01$, $0.02$, and $0.05$
are represented by $\Lambda_1$, $\Lambda_2$, and $\Lambda_3$, respectively.

For the N-BAR domain of amphiphysin 1,
the experimental data from Ref.~\citenum{tsai21} are used.
We assume that the average densities for $n_{\mathrm{v}}<50$\,$\mu$m$^{-2}$, 
$50$\,$\mu$m$^{-2}<n_{\mathrm{v}}<120$\,$\mu$m$^{-2}$, and $120$\,$\mu$m$^{-2}<n_{\mathrm{v}}<500$\,$\mu$m$^{-2}$
are $\phi_{\mathrm{v}}=0.0013$, $0.0043$, and $0.016$,  respectively,
where $n_{\mathrm{v}}$ represents the number density of proteins in the spherical vesicle region ($\phi_{\mathrm{v}}= n_{\mathrm{v}}a_{\mathrm {p}}$).
The fit deviations for $\phi_{\mathrm{v}}=0.0013$ and $0.0043$
are represented by $\Lambda_1$ and $\Lambda_2$, respectively.
The data of higher densities, i.e., $120$\,$\mu$m$^{-2}<n_{\mathrm{v}}<500$\,$\mu$m$^{-2}$,
are not used for the fitting, because they are widely distributed from $\phi_{\mathrm {cy}}/\phi_{\mathrm {v}} \simeq 3$ to $20$ for a narrow range of the tether curvature ($0.04{\mathrm{\,nm}}^{-1} \lesssim 1/R_{\mathrm {cy}} \lesssim 0.11{\mathrm{\,nm}}^{-1}$).
We compare only the mean value; when the density ratio at $1/R_{\mathrm {cy}}=0.07{\mathrm{\,nm}}^{-1}$ is in the range  $10 \lesssim\phi_{\mathrm {cy}}/\phi_{\mathrm {v}} \lesssim 15$,
we consider the fit to be good for $\phi_{\mathrm{v}}=0.016$.

\begin{figure}[tbh]
\includegraphics[]{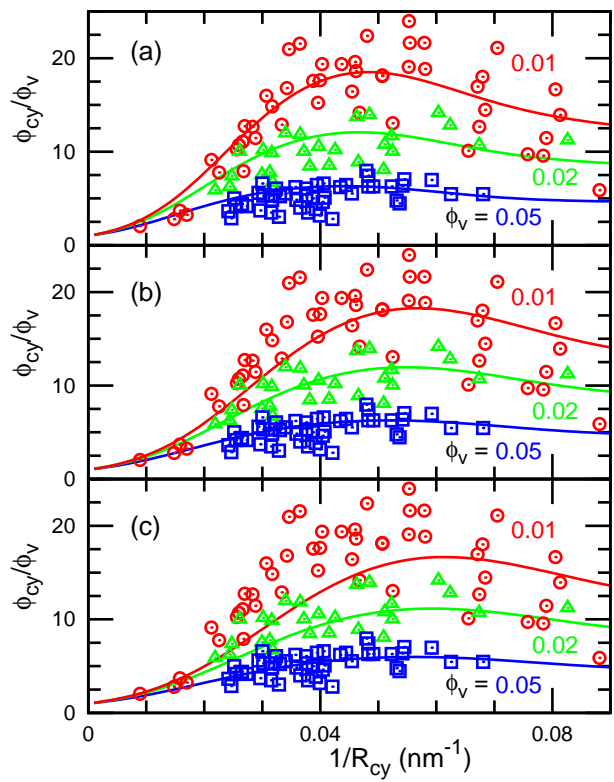}
\caption{
Fitting of the density--curvature curves for I-BAR-domain-binding based on the anisotropic protein model.
Circles, triangles, and squares indicate 
the experimental data of  $\phi_{\mathrm{cy}}/\phi_{\mathrm{v}}$ for $\phi_{\mathrm{v}}=0.01$, $0.02$, and $0.05$, respectively (reproduced from Ref.~\citenum{prev15}).
The solid lines are given by the theoretical results with the fitting parameters $\kappa_{\mathrm{p}}$ and $C_{\mathrm{p}}$ at $\kappa_{\mathrm{side}}=0$;
from top to bottom, $\phi_{\mathrm{v}}=0.01$, $0.02$, and $0.05$.
(a) Data at $\phi_{\mathrm{v}}=0.01$ are fitted:
 $\kappa_{\mathrm{p}}/k_{\mathrm{B}}T=100$ and $C_{\mathrm{p}}=0.043$\,nm$^{-1}$.
(b) Data at $\phi_{\mathrm{v}}=0.02$ are fitted:
 $\kappa_{\mathrm{p}}/k_{\mathrm{B}}T=72$ and $C_{\mathrm{p}}=0.0505$\,nm$^{-1}$.
(c) Data at $\phi_{\mathrm{v}}=0.05$ are fitted:
 $\kappa_{\mathrm{p}}/k_{\mathrm{B}}T=60$ and $C_{\mathrm{p}}=0.054$\,nm$^{-1}$.
}
\label{fig:iphiks0}
\end{figure}

\begin{figure}[tbh]
\includegraphics[]{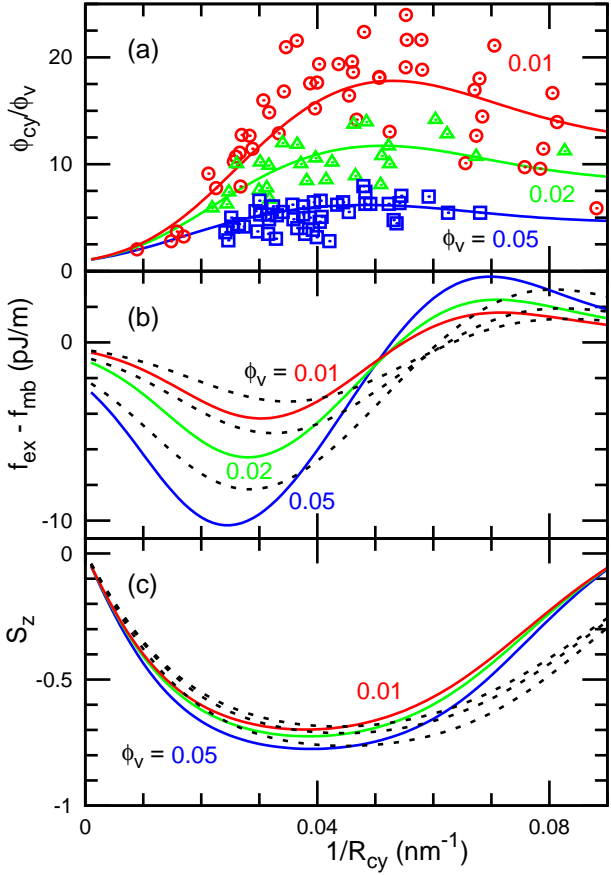}
\caption{
Fitting of the density--curvature curves for I-BAR-domain binding based on the anisotropic protein model.
The solid lines are given by the theoretical results at $\kappa_{\mathrm{p}}/k_{\mathrm{B}}T=82$ and $C_{\mathrm{p}}=0.047$\,nm$^{-1}$  
to minimize $\Lambda_1+\Lambda_2+\Lambda_3$.
(a) Circles, triangles, and squares indicate
the experimental data of $\phi_{\mathrm{cy}}/\phi_{\mathrm{v}}$ for $\phi_{\mathrm{v}}=0.01$, $0.02$, and $0.05$, respectively (reproduced from Ref.~\citenum{prev15}).
(b) Force generated by the protein binding.
(c) Degree $S_z$ of protein order along the membrane tube ($z$) axis.
(a, c) From top to bottom, $\phi_{\mathrm{v}}=0.01$, $0.02$, and $0.05$.
(b) From top to bottom, $\phi_{\mathrm{v}}=0.01$, $0.02$, and $0.05$ at $1/R_{\mathrm{cy}}< 0.05$\,nm$^{-1}$.
The dashed lines in (b) and (c) represent the data for $\kappa_{\mathrm{p}}/k_{\mathrm{B}}T=60$ and $C_{\mathrm{p}}=0.054$\,nm$^{-1}$
(the minimum value of $\Lambda_3$ corresponding to Fig.~\ref{fig:iphiks0}(c)).
}
\label{fig:iphi4}
\end{figure}

\begin{figure}[tbh]
\includegraphics[]{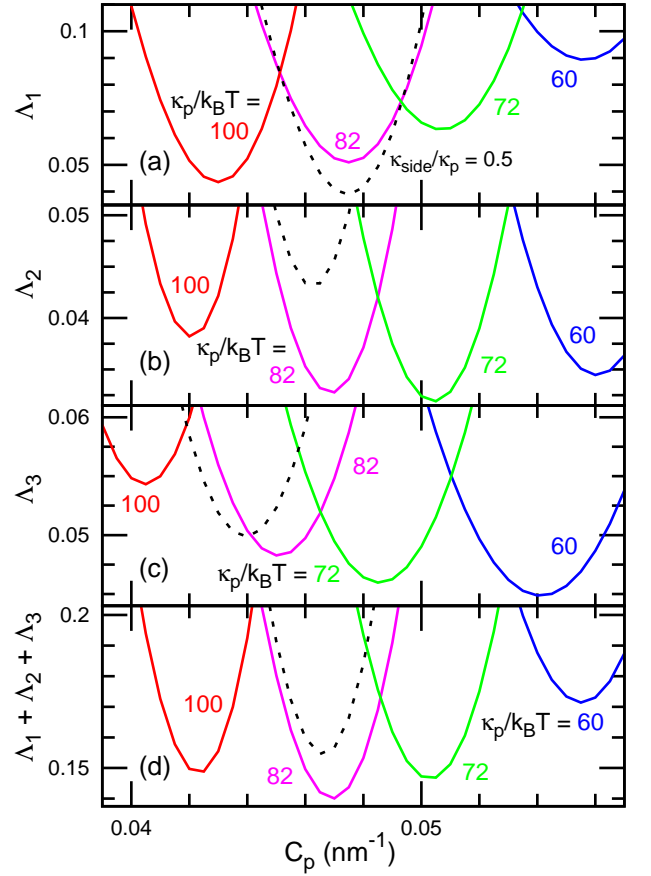}
\caption{
Fit deviation as a function of the protein curvature $C_{\mathrm{p}}$ for I-BAR-domain binding based on the anisotropic protein model.
The solid lines represent the theoretical results for $\kappa_{\mathrm{p}}/k_{\mathrm{B}}T=100$, $82$, $72$, and $60$ at $\kappa_{\mathrm{side}}=0$,
from left to right.
The dashed lines represent the data for $\kappa_{\mathrm{p}}/k_{\mathrm{B}}T=86$, $\kappa_{\mathrm{side}}/\kappa_{\mathrm{p}}=0.5$, and $C_{\mathrm{side}}=0$.
(a) Fit deviation $\Lambda_1$ for $\phi_{\mathrm{v}}=0.01$.
For $\kappa_{\mathrm{side}}=0$, the minimum value of $\Lambda_1^{\mathrm{min}}=0.044$ is obtained at $\kappa_{\mathrm{p}}/k_{\mathrm{B}}T=100$ and $C_{\mathrm{p}}=0.043$\,nm$^{-1}$ (corresponding to Fig.~\ref{fig:iphiks0}(a)).
For $\kappa_{\mathrm{side}}\ne 0$, a lower minimum $\Lambda_1^{\mathrm{min}}=0.039$ is obtained at $\kappa_{\mathrm{p}}/k_{\mathrm{B}}T=86$, $C_{\mathrm{p}}=0.0475$\,nm$^{-1}$, and $\kappa_{\mathrm{side}}/\kappa_{\mathrm{p}}=0.5$ (corresponding in Fig.~\ref{fig:iphiside}).
(b) Fit deviation $\Lambda_2$ for $\phi_{\mathrm{v}}=0.02$.
The minimum $\Lambda_2^{\mathrm{min}}=0.032$ is obtained at $\kappa_{\mathrm{p}}/k_{\mathrm{B}}T=72$, $C_{\mathrm{p}}=0.0505$\,nm$^{-1}$, and $\kappa_{\mathrm{side}}=0$ (corresponding to Fig.~\ref{fig:iphiks0}(b)).
(c) Fit deviation $\Lambda_3$ for $\phi_{\mathrm{v}}=0.05$.
The minimum $\Lambda_3^{\mathrm{min}}=0.045$ is obtained at $\kappa_{\mathrm{p}}/k_{\mathrm{B}}T=60$, $C_{\mathrm{p}}=0.054$\,nm$^{-1}$, and $\kappa_{\mathrm{side}}=0$ (corresponding to Fig.~\ref{fig:iphiks0}(c)).
(d) Sum of fit deviations $\Lambda_1+\Lambda_2+\Lambda_3$.
The minimum value $(\Lambda_1+\Lambda_2+\Lambda_3)^{\mathrm{min}}=0.140$ is obtained at $\kappa_{\mathrm{p}}/k_{\mathrm{B}}T=82$, $C_{\mathrm{p}}=0.047$\,nm$^{-1}$, and $\kappa_{\mathrm{side}}=0$ (corresponding to Fig.~\ref{fig:iphi4}).
}
\label{fig:iSks0}
\end{figure}

\begin{figure}[tbh]
\includegraphics[]{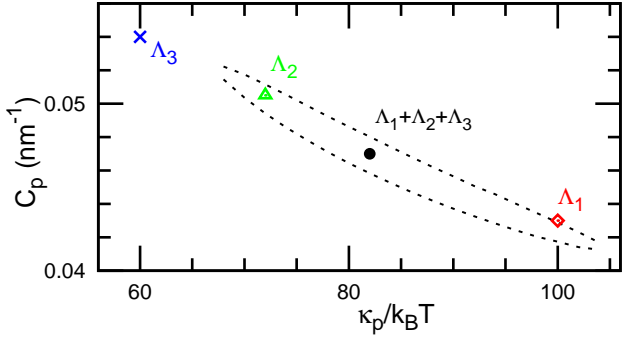}
\caption{
Two-dimensional map of the fit deviations of $\Lambda_1+\Lambda_2+\Lambda_3$ for I-BAR-domain binding based on the anisotropic protein model.
The circle represents the minimum point, and the dashed lines represent the contour of values exceeding the minimum by $10$\%.
The diamond, triangle, and cross indicate the minimum points for $\Lambda_1$,  $\Lambda_2$, and  $\Lambda_3$, respectively.
}
\label{fig:iamin}
\end{figure}

\begin{figure}[tbh]
\includegraphics[]{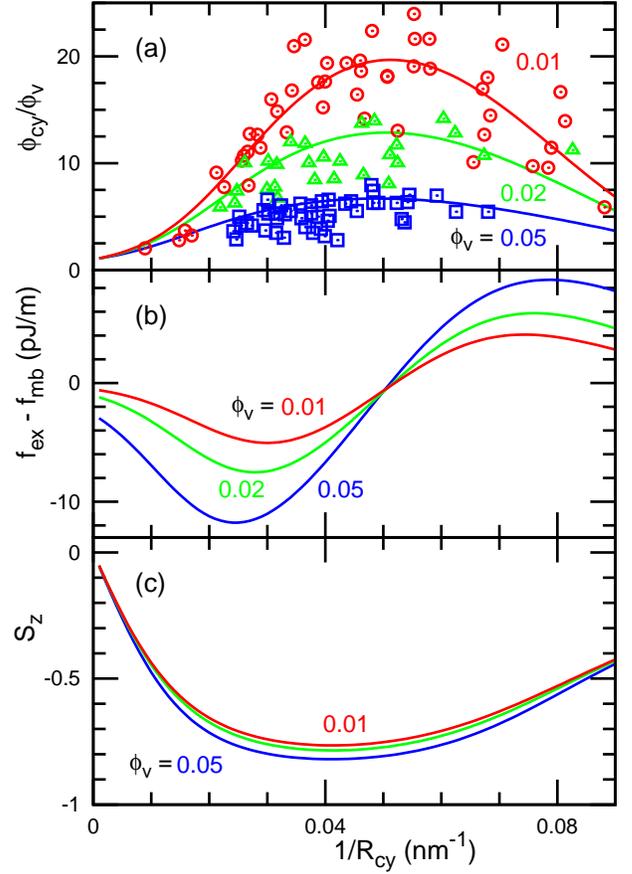}
\caption{
Fitting of the density--curvature curves for I-BAR-domain binding based on the anisotropic protein model.
The solid lines are given by the theoretical results at $\kappa_{\mathrm{p}}/k_{\mathrm{B}}T=86$, $C_{\mathrm{p}}=0.0475$\,nm$^{-1}$, $\kappa_{\mathrm{side}}/\kappa_{\mathrm{p}}=0.5$, and $C_{\mathrm{side}}=0$
to minimize $\Lambda_1$.
(a) Circles, triangles, and squares indicate
the experimental data of $\phi_{\mathrm{cy}}/\phi_{\mathrm{v}}$ for $\phi_{\mathrm{v}}=0.01$, $0.02$, and $0.05$, respectively (reproduced from Ref.~\citenum{prev15}).
(b) Force generated by the protein binding.
(c) Degree $S_z$ of protein order along the membrane tube ($z$) axis.
(a, c) From top to bottom, $\phi_{\mathrm{v}}=0.01$, $0.02$, and $0.05$.
(b) From top to bottom, $\phi_{\mathrm{v}}=0.01$, $0.02$, and $0.05$ at $1/R_{\mathrm{cy}}< 0.05$\,nm$^{-1}$.
}
\label{fig:iphiside}
\end{figure}

\begin{figure}[tbh]
\includegraphics[]{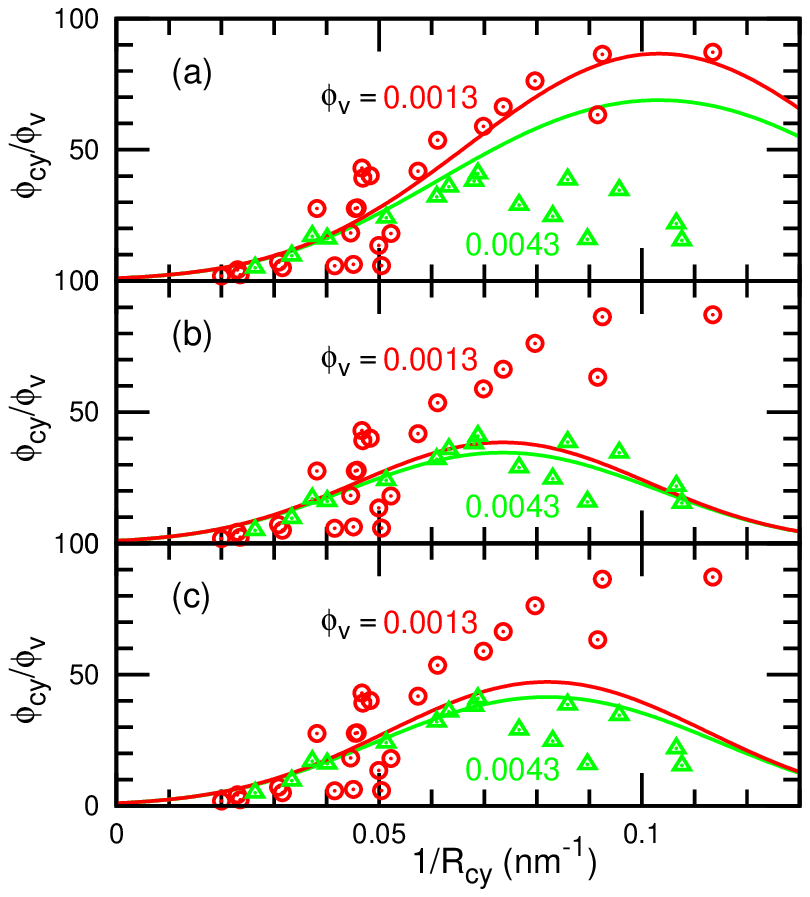}
\caption{
Fitting of the density--curvature curves for N-BAR-domain binding based on the isotropic protein model.
Circles and triangles indicate
the experimental data of $\phi_{\mathrm{cy}}/\phi_{\mathrm{v}}$ for $\phi_{\mathrm{v}}=0.0013$ and $0.0043$, respectively (reproduced from Ref.~\citenum{tsai21} with permission from the Royal Society of Chemistry).
The solid lines are given by eqn~(\ref{eq:phiia}) with the fitting parameters $\kappa_{\mathrm{dif}}$ and $C_{\mathrm{s}}$;
from top to bottom, $\phi_{\mathrm{v}}=0.0013$ and $0.0043$.
(a) Data at $\phi_{\mathrm{v}}=0.0013$ are fitted:
 $\kappa_{\mathrm{dif}}/k_{\mathrm{B}}T=17.2$ and $C_{\mathrm{s}}=0.1032$\,nm$^{-1}$.
(b) Data at $\phi_{\mathrm{v}}=0.0043$ are fitted:
 $\kappa_{\mathrm{dif}}/k_{\mathrm{B}}T=27.4$ and $C_{\mathrm{s}}=0.0735$\,nm$^{-1}$.
(c) Both datasets are fitted:
 $\kappa_{\mathrm{dif}}/k_{\mathrm{B}}T=23.3$ and $C_{\mathrm{s}}=0.0820$\,nm$^{-1}$.
}
\label{fig:niphi}
\end{figure}

\begin{figure}[tbh]
\includegraphics[]{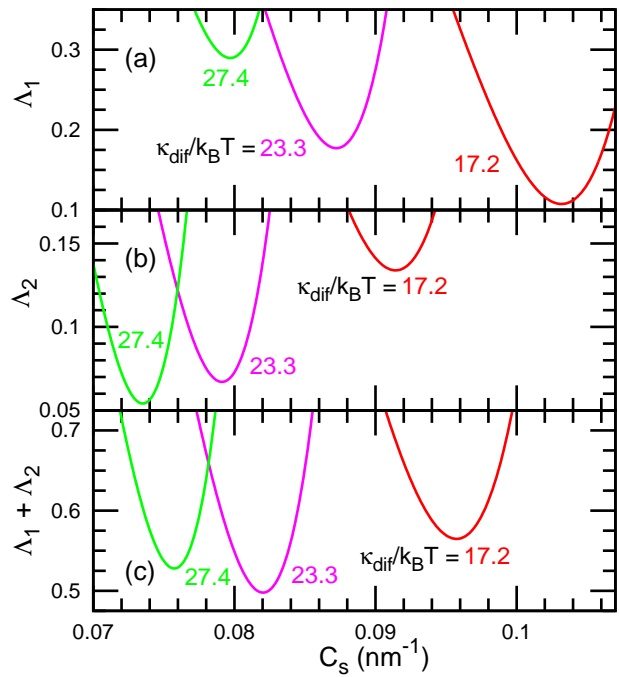}
\caption{
Fit deviation as a function of the sensing curvature $C_{\mathrm{s}}$ for N-BAR-domain binding based on the isotropic protein model.
From left to right,  $\kappa_{\mathrm{dif}}/k_{\mathrm{B}}T=27.4$, $23.3$, and $17.2$.
(a) Fit deviation $\Lambda_1$ for $\phi_{\mathrm{v}}=0.0013$.
The minimum $\Lambda_1^{\mathrm{min}}=0.1075$ is obtained at $\kappa_{\mathrm{dif}}/k_{\mathrm{B}}T=17.2$ and $C_{\mathrm{s}}=0.1032$\,nm$^{-1}$ (corresponding to Fig.~\ref{fig:niphi}(a)).
(b) Fit deviation $\Lambda_2$ for $\phi_{\mathrm{v}}=0.0043$.
The minimum $\Lambda_2^{\mathrm{min}}=0.0541$ is obtained at $\kappa_{\mathrm{dif}}/k_{\mathrm{B}}T=27.4$ and $C_{\mathrm{s}}=0.0735$\,nm$^{-1}$ (corresponding to Fig.~\ref{fig:niphi}(b)).
(c) Sum of fit deviations $\Lambda_1+\Lambda_2$.
The minimum $(\Lambda_1+\Lambda_2)^{\mathrm{min}}=0.498$ is obtained at $\kappa_{\mathrm{dif}}/k_{\mathrm{B}}T=23.3$ and $C_{\mathrm{s}}=0.0820$\,nm$^{-1}$ (corresponding to Fig.~\ref{fig:niphi}(c)).
}
\label{fig:niS}
\end{figure}

\begin{figure}[tbh]
\includegraphics[]{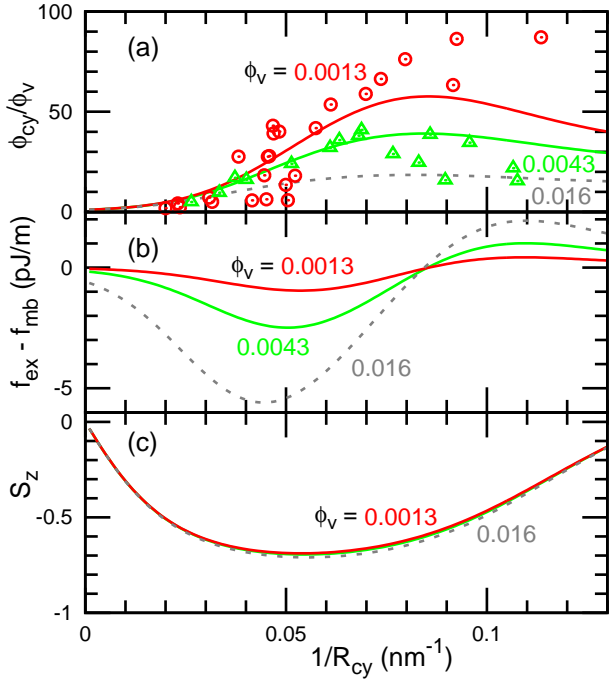}
\caption{
Fitting of the density--curvature curves for N-BAR-domain binding based on the anisotropic protein model.
The lines are given by the theoretical results at $\kappa_{\mathrm{p}}/k_{\mathrm{B}}T=39$, $C_{\mathrm{p}}=0.072$\,nm$^{-1}$, and $\kappa_{\mathrm{side}}=0$ for minimizing $\Lambda_1+\Lambda_2$.
(a) 
Circles and triangles indicate
the experimental data of $\phi_{\mathrm{cy}}/\phi_{\mathrm{v}}$ for $\phi_{\mathrm{v}}=0.0013$ and $0.0043$, respectively (reproduced from Ref.~\citenum{tsai21} with permission from the Royal Society of Chemistry).
(b) Force generated by the protein binding.
(c) Degree $S_z$ of protein order along the membrane tube ($z$) axis.
(a, c) From top to bottom, $\phi_{\mathrm{v}}=0.0013$, $0.0043$, and $0.016$.
(b) From top to bottom, $\phi_{\mathrm{v}}=0.0013$, $0.0043$, and $0.016$ at $1/R_{\mathrm{cy}} < 0.08$\,nm$^{-1}$.
}
\label{fig:nphiks3}
\end{figure}

\begin{figure}[tbh]
\includegraphics[]{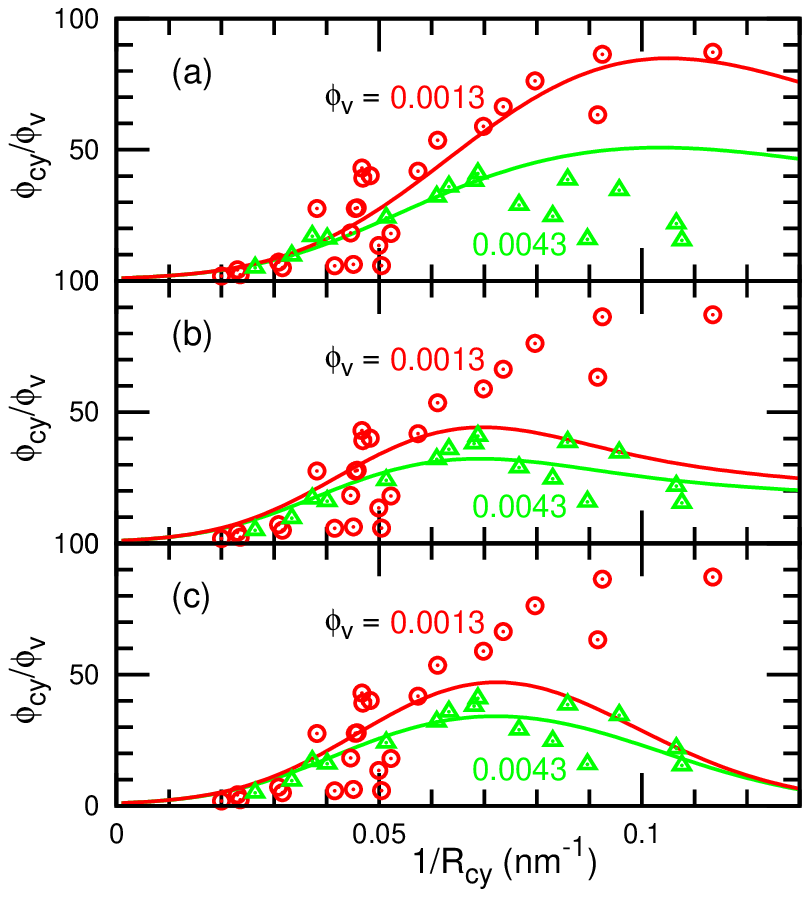}
\caption{
Fitting of the density--curvature curves for N-BAR-domain binding based on the anisotropic protein model.
Circles and triangles indicate
the experimental data of $\phi_{\mathrm{cy}}/\phi_{\mathrm{v}}$ for $\phi_{\mathrm{v}}=0.0013$ and $0.0043$, respectively (reproduced from Ref.~\citenum{tsai21} with permission from the Royal Society of Chemistry).
The solid lines are given by the theoretical results;
from top to bottom, $\phi_{\mathrm{v}}=0.0013$ and $0.0043$.
(a) Data at $\phi_{\mathrm{v}}=0.0013$ are fitted:
 $\kappa_{\mathrm{p}}/k_{\mathrm{B}}T=28$ and $C_{\mathrm{p}}=0.0895$\,nm$^{-1}$ at $\kappa_{\mathrm{side}}=0$.
(b) Data at $\phi_{\mathrm{v}}=0.0043$ are fitted:
 $\kappa_{\mathrm{p}}/k_{\mathrm{B}}T=55$ and $C_{\mathrm{p}}=0.0585$\,nm$^{-1}$ at $\kappa_{\mathrm{side}}=0$.
(c) Data at $\phi_{\mathrm{v}}=0.0043$ are fitted:
 $\kappa_{\mathrm{p}}/k_{\mathrm{B}}T=46$, $C_{\mathrm{p}}=0.066$\,nm$^{-1}$, $\kappa_{\mathrm{side}}=0.5\kappa_{\mathrm{p}}$, and $C_{\mathrm{side}}=-0.0025$\,nm$^{-1}$.
}
\label{fig:nphia}
\end{figure}

\begin{figure}[tbh]
\includegraphics[]{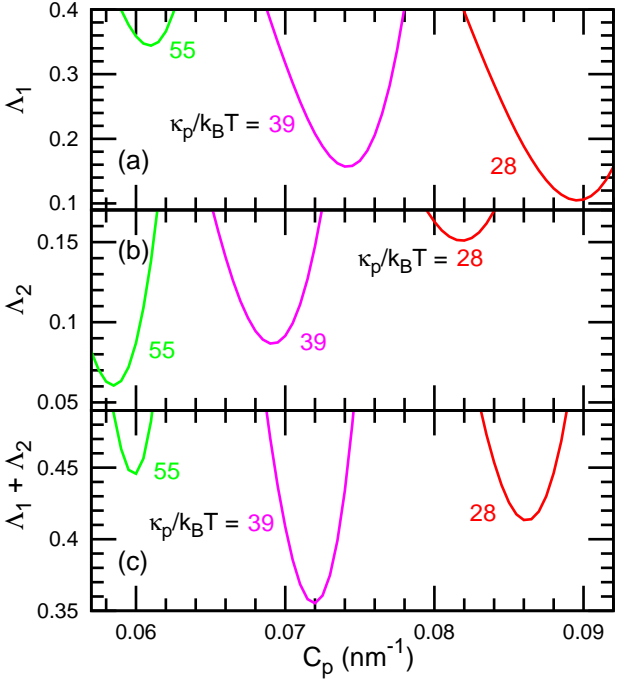}
\caption{
Fit deviation as a function of the protein curvature $C_{\mathrm{p}}$ for N-BAR-domain binding based on the anisotropic protein model at $\kappa_{\mathrm{side}}=0$.
From left to right,  $\kappa_{\mathrm{p}}/k_{\mathrm{B}}T=55$, $39$, and $28$.
(a) Fit deviation $\Lambda_1$ for $\phi_{\mathrm{v}}=0.0013$.
The minimum $\Lambda_1^{\mathrm{min}}=0.105$ is obtained at $\kappa_{\mathrm{p}}/k_{\mathrm{B}}T=28$ and $C_{\mathrm{p}}=0.0895$\,nm$^{-1}$ (corresponding to Fig.~\ref{fig:nphia}(a)).
(b) Fit deviation $\Lambda_2$ for $\phi_{\mathrm{v}}=0.0043$.
The minimum $\Lambda_2^{\mathrm{min}}=0.061$ is obtained at $\kappa_{\mathrm{p}}/k_{\mathrm{B}}T=55$ and $C_{\mathrm{p}}=0.0585$\,nm$^{-1}$ (corresponding to Fig.~\ref{fig:nphia}(b)).
(c) Sum of fit deviations $\Lambda_1+\Lambda_2$.
The minimum $(\Lambda_1+\Lambda_2)^{\mathrm{min}}=0.355$ is obtained at $\kappa_{\mathrm{p}}/k_{\mathrm{B}}T=39$ and $C_{\mathrm{p}}=0.072$\,nm$^{-1}$ (corresponding to Fig.~\ref{fig:nphiks3}).
}
\label{fig:nSks0}
\end{figure}

\begin{figure}[tbh]
\includegraphics[]{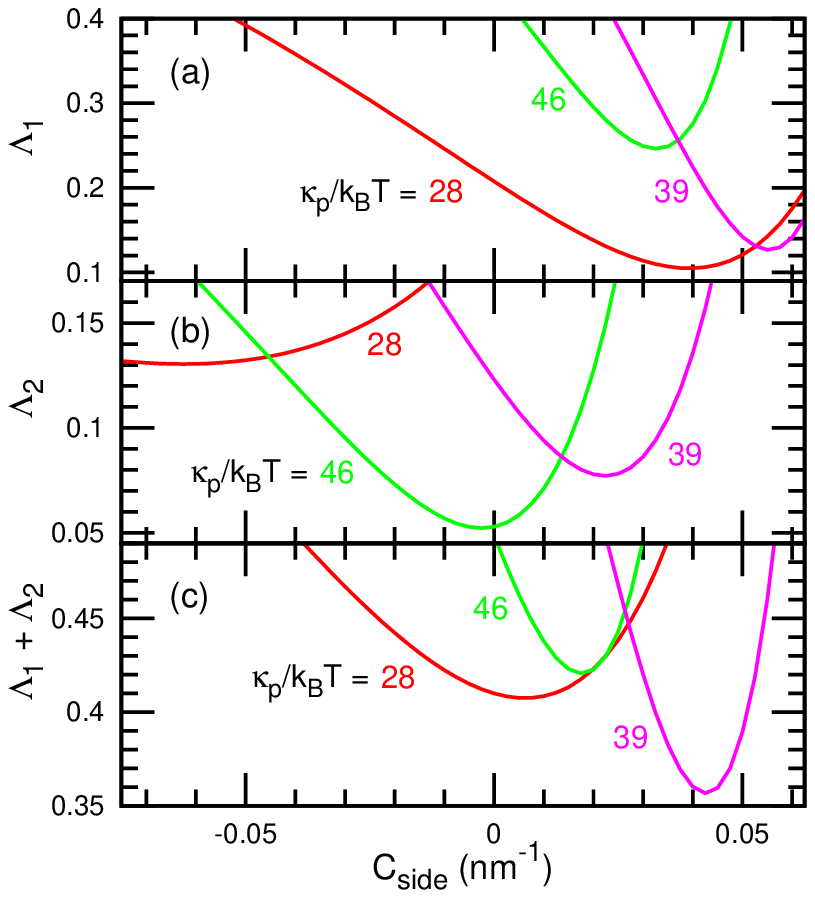}
\caption{
Fit deviation as a function of the side curvature $C_{\mathrm{side}}$ for N-BAR-domain binding based on the anisotropic protein model at $\kappa_{\mathrm{side}}>0$.
From left to right,  $(\kappa_{\mathrm{p}}/k_{\mathrm{B}}T, C_{\mathrm{p}}$ (nm$^{-1}$)$)=(28, 0.087)$, $(46, 0.066)$, and $(39, 0.068)$ with $\kappa_{\mathrm{side}}/\kappa_{\mathrm{p}}=0.5$.
(a) Fit deviation $\Lambda_1$ for $\phi_{\mathrm{v}}=0.0013$.
The minimum $\Lambda_1^{\mathrm{min}}=0.105$ is obtained at $(\kappa_{\mathrm{p}}/k_{\mathrm{B}}T, C_{\mathrm{p}}$ (nm$^{-1}$)$, C_{\mathrm{side}}$ (nm$^{-1}$)$, \kappa_{\mathrm{side}}/\kappa_{\mathrm{p}})=(28, 0.087, 0.04, 0.5)$.
(b) Fit deviation $\Lambda_2$ for $\phi_{\mathrm{v}}=0.0043$.
The minimum $\Lambda_2^{\mathrm{min}}=0.052$ is obtained at $(\kappa_{\mathrm{p}}/k_{\mathrm{B}}T, C_{\mathrm{p}}$ (nm$^{-1}$)$, C_{\mathrm{side}}$ (nm$^{-1}$)$, \kappa_{\mathrm{side}}/\kappa_{\mathrm{p}})=(46, 0.066, -0.0025, 0.5)$ (Fig.~\ref{fig:nphia}(c)).
(c) Sum of fit deviations $\Lambda_1+\Lambda_2$.
The minimum $(\Lambda_1+\Lambda_2)^{\mathrm{min}}=0.357$ is obtained at $(\kappa_{\mathrm{p}}/k_{\mathrm{B}}T, C_{\mathrm{p}}$ (nm$^{-1}$)$, C_{\mathrm{side}}$ (nm$^{-1}$)$, \kappa_{\mathrm{side}}/\kappa_{\mathrm{p}})=(39, 0.068, 0.0425, 0.5)$.
}
\label{fig:nSside}
\end{figure}

\section{Binding of I-BAR domains}\label{sec:ibar0}

\subsection{Isotropic protein model}\label{sec:ibari}

Before examining the anisotropic protein model,
 we fit the experimental data of I-BAR domains using the isotropic protein model with eqn~(\ref{eq:phiia}).
The rigidity difference $\kappa_{\mathrm{dif}}$ and sensing curvature $C_{\mathrm{s}}$ are fitted.
Figure~\ref{fig:iiphi}(a), (b), and (c) show the best fits for  $\phi_{\mathrm{v}}=0.01$, $0.02$, and $0.05$,
to minimize $\Lambda_1$, $\Lambda_2$, and $\Lambda_3$, respectively (see Figs.~\ref{fig:iimin} and \ref{fig:iiS}).
The target density data are fitted very well but the other two datasets exhibit large deviations.
When the sum $\Lambda_1+\Lambda_2+\Lambda_3$ is minimized, the obtained curves are close to the results of the fit to
the middle data, i.e., $\Lambda_2$ (compare Fig.~\ref{fig:iiphi}(d) and (b)).
Therefore, not all of the data can be reproduced together using the isotropic protein model.
Note that each dataset was fit separately (i.e., different values of the fitting parameters) in Ref.~\citenum{prev15}.

As $\kappa_{\mathrm{dif}}$ increases, the lowest value of $\Lambda$ is obtained 
at lower $C_{\mathrm{s}}$ (see Fig.~\ref{fig:iiS}).
This is because $\kappa_{\mathrm{dif}}C_{\mathrm{s}}$ ($=\kappa_{\mathrm{pi}}C_0$) is 
a factor of the major term in the exponent of eqn~(\ref{eq:phiia}).
Thus, $\Lambda$ values close to the minimum are obtained in the long narrow regions 
along the solid curves in Fig.~\ref{fig:iimin}.
When a statistical error is considered to be $10$\% of $\Lambda$,
$\kappa_{\mathrm{dif}}$ and $C_{\mathrm{s}}$ are in the regions of the long ellipses in Fig.~\ref{fig:iimin}.
The anisotropic model exhibits a similar dependence for $\kappa_{\mathrm{p}}$ and $C_{\mathrm{p}}$ as described later.

Equation~(\ref{eq:phiis}) for $\phi_{\mathrm{cy}}\ll 1$ has been used
instead of eqn~(\ref{eq:phiia}) in the previous studies~\cite{tsai21,prev15,rosh17,yang22}.
We compared the results of these two equations for $\Lambda_1$, $\Lambda_2$, and $\Lambda_3$ 
using  the best-fit parameters at $\phi_{\mathrm{v}}=0.01$, $0.02$, and $0.05$, respectively.
We found that the low-density limit approximation overestimates the values of $\phi_{\mathrm{cy}}$ by approximately $20$\%, $30$\%, and $40$\%, respectively.
A bound protein prevents the binding of other proteins to the same position
(the binding rate is proportional to $(1-\phi)$)~\cite{gout21}.
Although this effect is negligible in the limit $\phi_{\mathrm{cy}}\ll 1$,
it becomes recognizable at $\phi_{\mathrm{cy}} \simeq 0.1$
(compare functions $\exp(-x)$ and  $1/[1+\exp(x)]$).
Therefore, eqn~(\ref{eq:phiia}) should be used in $\phi_{\mathrm{cy}} \gtrsim 0.1$.

\subsection{Anisotropic protein model}\label{sec:ibara}

The experimental data of the I-BAR domains are fitted using the anisotropic protein model.
First, we use the protein rigidity $\kappa_{\mathrm{p}}$ and protein curvature $C_{\mathrm{p}}$ as the fitting parameters with $\kappa_{\mathrm{side}}=0$,
as shown in Figs.~\ref{fig:iphiks0}--\ref{fig:iamin}.
To determine the minimum for each fit, the parameters are varied discretely with $\Delta\kappa_{\mathrm{p}}= 2k_{\mathrm{B}}T$ and $\Delta C_{\mathrm{p}}=0.0005$\,nm$^{-1}$.
Very good agreement is obtained for not only the target density--curvature curve but also the other two curves (see Fig.~\ref{fig:iphiks0}).
The minimum $\Lambda$ values for the target curves are almost identical to those of the isotropic protein model 
(compare Fig.~\ref{fig:iSks0}(a)--(c) and Fig.~\ref{fig:iiS}(a)--(c)).
However, the sum, $\Lambda_1+\Lambda_2+\Lambda_3$, is significantly reduced;
thus, the total fitness is improved (compare Fig.~\ref{fig:iSks0}(d) and Fig.~\ref{fig:iiS}(d)).

Although the density--curvature curves are sufficiently reproduced using $\kappa_{\mathrm{side}}=0$,
a deviation is  recognized in $1/R_{\mathrm{cy}} \gtrsim 0.07$\,nm$^{-1}$ for $\phi_{\mathrm{v}}=0.01$.
In the  experimental data, 
the protein density decreases by a larger amount as $1/R_{\mathrm{cy}}$ increases.
The other two curves do not exhibit such a deviation,
possibly owing to limited data for the narrow tubes 
(one and no data points for $1/R_{\mathrm{cy}}> 0.07$\,nm$^{-1}$ at  $\phi_{\mathrm{v}}=0.02$ and $0.05$, respectively).
This deviation can be eliminated by using a finite value of the bending rigidity $\kappa_{\mathrm{side}}$ in the side direction.
In addition,  $\kappa_{\mathrm{side}}$ is varied discretely with $\Delta\kappa_{\mathrm{side}}= 0.5\kappa_{\mathrm{p}}$ at $C_{\mathrm{side}}=0$ for $0\le\kappa_{\mathrm{side}}\le\kappa_{\mathrm{p}}$.
A better fit is obtained for $\phi_{\mathrm{v}}=0.01$, as shown in Figs.~\ref{fig:iSks0}(a) and \ref{fig:iphiside}.
Although $C_{\mathrm{side}}$ does not vary together here,
$\Lambda_1$ has a minimum at $C_{\mathrm{side}}=0$ for the variation in $C_{\mathrm{side}}$ with $\Delta C_{\mathrm{side}} =0.0025$\,nm$^{-1}$
 when the other parameters are fixed.
Thus, the zero side curvature is reasonable.
In contrast to  $\Lambda_1$, lower values of the others ($\Lambda_2$, $\Lambda_3$, and $\Lambda_1+\Lambda_2+\Lambda_3$) 
are not obtained using $\kappa_{\mathrm{side}}\ne 0$.
Thus, the best fit for the total data ($(\Lambda_1+\Lambda_2+\Lambda_3)^{\mathrm{min}}=0.140$) is given at $\kappa_{\mathrm{side}}=0$, as shown in Fig.~\ref{fig:iphi4}.

When a $10$\% larger value of $\Lambda_1+\Lambda_2+\Lambda_3$ is allowed as a statistical error,
the elliptical region surrounded by dashed lines in Fig.~\ref{fig:iamin} is the expected range of $\kappa_{\mathrm{p}}$ and $C_{\mathrm{p}}$.
The minimum points of $\Lambda_1$ and $\Lambda_2$ are also included in this range.
Hence, we concluded that the I-BAR domain has $\kappa_{\mathrm{p}}/k_{\mathrm{B}}T= 82\pm 20$ 
and $C_{\mathrm{p}}(\mathrm{nm}^{-1}) = 0.047 - 0.0003(\kappa_{\mathrm{p}}/k_{\mathrm{B}}T-82) \pm 0.001$.

Although the experimental data are well-fitted without the side rigidity,
the existence of the side rigidity is not excluded.
The measurement of other quantities can increase the estimation accuracy
for the mechanical properties of proteins.
Here, we propose the force $f_{\mathrm{ex}}$ and orientational degree $S_z$ along the tube axis
as candidates. 
The force $f_{\mathrm{ex}}$ has been experimentally measured from
the position of optically trapped beads~\cite{heir96,dimo14}.
Force modification due to the binding of BAR domains has been reported~\cite{prev15,sorr12}.
Although $S_z$ has not been experimentally measured,
it is measurable using polarizers in principle.
As the tube curvature $1/R_{\mathrm{cy}}$ increases,
the two quantities vary in different manners with respect to the protein density,
as shown in Figs.~\ref{fig:iphi4} and \ref{fig:iphiside}.
Each has a minimum value at a tube curvature lower than the sensing curvature.
Since the densities are sufficiently low, $S_z$ exhibits only a small dependence on $\phi_{\mathrm{v}}$.
Importantly, they vary with changes in the bending parameters,
although the density curves do not vary significantly
(compare the solid and dashed lines in Fig.~\ref{fig:iphi4}(b) and (c),
and also see Fig.~\ref{fig:iphiside}(b) and (c)).
This suggests that the bending rigidity and curvature of proteins can be more accurately estimated
through additional fitting of $f_{\mathrm{ex}}$ and $S_z$ to the experimental data.

\section{Binding of N-BAR domains}\label{sec:nbar0}

\subsection{Isotropic protein model}\label{sec:nbari}

In this section, 
we consider the binding of N-BAR domains reported in Refs.~\citenum{sorr12} and \citenum{tsai21}.
First, we examine the isotropic protein model, as for the I-BAR domains considered  in Sec.~\ref{sec:ibar0}.
Figures \ref{fig:niphi} and \ref{fig:niS} show the fitted density--curvature curves and fit deviations, respectively,
using eqn~(\ref{eq:phiia}).
The target curve is well-fitted, whereas the other is not,
 similar to the case of the I-BAR domain.

\subsection{Anisotropic protein model}\label{sec:nbara}

The experimental data of the N-BAR domains are fitted using the anisotropic protein model.
First, the fitting is performed using the fitting parameters $\kappa_{\mathrm{p}}$ and $C_{\mathrm{p}}$ at $\kappa_{\mathrm{side}}=0$.
The parameters are varied discretely with $\Delta\kappa_{\mathrm{p}}= k_{\mathrm{B}}T$ and $\Delta C_{\mathrm{p}}=0.0005$\,nm$^{-1}$.
Figure \ref{fig:nphiks3}(a) shows the fitted density--curvature curves for
minimizing $\Lambda_1+\Lambda_2$.
The two curves exhibit far better agreements than those obtained using the isotropic protein model (compare Figs.~\ref{fig:nphiks3}(a) and \ref{fig:niphi}(c)).
Nonetheless, the deviations from the curves fitted to the data for $\phi_{\mathrm {v}}=0.0013$ or $0.0043$ 
 are large.
Additionally, the curve obtained for $\phi_{\mathrm {v}}=0.016$ is slightly exceeded the expected values 
($\phi_{\mathrm {cy}}/\phi_{\mathrm {v}} = 18$ at $1/R_{\mathrm {cy}}=0.07$\,nm$^{-1}$, as indicated by the dashed line in Fig.~\ref{fig:nphiks3}(a)).

The fit deviation of the curve minimizing $\Lambda_1$ is identical to that of the isotropic model, 
but that for $\Lambda_2$ is slightly worse (see Figs.~\ref{fig:nphia}(a), (b) and \ref{fig:nSks0}).
Similar to the case of the I-BAR domain with $\phi_{\mathrm {v}}=0.01$,
this is due to the smaller reduction in $\phi_{\mathrm {cy}}/\phi_{\mathrm {v}}$ at high tube curvatures (see Fig.~\ref{fig:nphia}(b)).
Thus, we perform the fit with a finite side rigidity for $0<\kappa_{\mathrm{side}}\le \kappa_{\mathrm{p}}$,
as shown in Figs.~\ref{fig:nphia}(c) and \ref{fig:nSside}.
The parameters are varied discretely with $\Delta\kappa_{\mathrm{p}}= k_{\mathrm{B}}T$, $\Delta C_{\mathrm{p}}=0.001$\,nm$^{-1}$,
 $\Delta \kappa_{\mathrm{side}}=0.5\kappa_{\mathrm{p}}$, and $\Delta C_{\mathrm{side}}=0.0025$\,nm$^{-1}$.
A better fit to the data at $\phi_{\mathrm {v}}=0.0043$ is obtained ($\Lambda_2^{\mathrm{min}}$ is 4\% smaller than that of the isotropic model).
For the other fit deviations ($\Lambda_1+\Lambda_2$ and $\Lambda_1$), the minimum values are almost identical to those at $\kappa_{\mathrm{side}}=0$.
Interestingly, in all cases, 
better fits are obtained at $\kappa_{\mathrm{side}}/\kappa_{\mathrm{p}}=0.5$ than at $\kappa_{\mathrm{side}}/\kappa_{\mathrm{p}}=1$.

Since the density $\phi_{\mathrm {v}}$ has a wide distribution in the experimental data,
we additionally performed the fitting at $\phi_{\mathrm{v}}=0.001$ and $0.0015$ 
to examine the effect of the choice of $\phi_{\mathrm{v}}$ values for $n_{\mathrm{v}}<50$\,$\mu$m$^{-2}$.
The minimum value of the sum $\Lambda_1+\Lambda_2$ decreases by 5\% and increases by 4\%
for  $\phi_{\mathrm{v}}=0.001$ and $0.0015$, ($(\Lambda_1+\Lambda_2)^{\mathrm{min}}=0.336$ and  $0.368$), respectively.
The corresponding values of $\kappa_{\mathrm{p}}/k_{\mathrm{B}}T$ and $C_{\mathrm{p}}$ (nm$^{-1}$)
are shifted by only $1$  and $0.001$, respectively.
For $\Lambda_1$, $\kappa_{\mathrm{p}}/k_{\mathrm{B}}T$ and $C_{\mathrm{p}}$ (nm$^{-1}$) are shifted by only $2$  and $0.0025$, respectively,
and the minimum value does not change.
Therefore, a slight variation in $\phi_{\mathrm{v}}$ does not result in a significant change.

Next, we consider the effects of variations in the other fixed parameters.
An aspect ratio of $d_{\mathrm {el}}=3$ is used for the N-BAR domains.
Since the protein densities are low, the excluded-volume effect is weak.
We examined the cases of $d_{\mathrm {el}}=2$ and $4$ with the best-fit conditions, 
and only obtained a 5\% decrease in $(\Lambda_1+\Lambda_2)^{\mathrm{min}}$ at $d_{\mathrm {el}}=4$.
Thus, we concluded that the N-BAR domains have $30\lesssim\kappa_{\mathrm{p}}/k_{\mathrm{B}}T \lesssim 60$ 
and $0.06 \lesssim C_{\mathrm{p}}(\mathrm{nm}^{-1})\lesssim 0.09$.

Note that the protein shape on the membrane surface might be modified by the membrane curvature.
However, as discussed above, their effects are considered small for N-BAR domains. 
For other proteins that exhibit large structural changes on a membrane surface as observed in deformable colloids~\cite{midy23},
their deformations can have significant effects.

The protein area $a_{\mathrm {p}}$ can be slightly varied by area definition.
In our previous studies~\cite{tozz21,roux21,nogu22},
we used $a_{\mathrm {p}}=60$\,nm$^2$, since the protein was approximated as elliptical with $\ell_1=15$\,nm and $\ell_2=5$\,nm,
and the protein area partially included bare membrane regions.
To fit a density--curvature curve,
 the area $a_{\mathrm {p}}$ only appears as pairs with bending rigidities in the theoretical models
($a_{\mathrm {p}}\kappa_{\mathrm{p}}$ and $a_{\mathrm {p}}\kappa_{\mathrm{side}}$ in the anisotropic model
and $a_{\mathrm {p}}\kappa_{\mathrm{dif}}$ in the isotropic model).
Thus, the area variation influences the bending rigidity in accordance with $\kappa'=\kappa a_{\mathrm {p}}/a_{\mathrm {p}}'$
(e.g., a $20$\% decrease in $a_{\mathrm {p}}$ results in a $25$\% increase in the bending rigidities).
The force, $f_{\mathrm {ex}}-f_{\mathrm {mb}}$, is also modified by the factor $a_{\mathrm {p}}/a_{\mathrm {p}}'$.
When the protein density is measured as the number density $n_{\mathrm{v}}$,
the area fraction is slightly modified according to the area definition ($\phi_{\mathrm{v}}=n_{\mathrm{v}}a_{\mathrm {p}}$).
However, this results in only slight changes, as discussed above.

In our previous study~\cite{nogu22},
we compared the results of the anisotropic model and a coarse-grained membrane simulation.
When the proteins are distributed homogeneously in the membrane,
the results agree very well.
In contrast, deviations are obtained, when the proteins form small clusters.
Since the clusters have a larger area than the individual proteins, they are more oriented in the preferred direction. 
The ratio of the clusters increases with the protein density.
Thus, we consider that the deviation in this study suggests non-negligible cluster formation
of the N-BAR domains.
Note that the clustering can be induced by direct attractive interaction between proteins 
and membrane-mediated interactions.
Large clusters can deform a membrane tube into polygonal shapes~\cite{nogu22a,nogu15b}.

\section{Summary}\label{sec:sum}

We have developed an estimation method for the mechanical properties of bound proteins based on the experiments of tethered vesicles
and applied it to the I-BAR and N-BAR domains.
When the anisotropy of the proteins is taken into account,
the experimental data are reproduced far better.
When the classical isotropic model is used, 
each density--curvature curve is well reproduced but the other curves largely deviate.
When the recently developed anisotropic model is used,
this deviation is significantly reduced.
For the I-BAR domains, all three curves are well-fitted by a single parameter set,
and the bending rigidity $\kappa_{\mathrm{p}}/k_{\mathrm{B}}T= 82$ 
and spontaneous curvature $C_{\mathrm{p}}(\mathrm{nm}^{-1}) = 0.047$ along the protein axis are determined.
The estimation errors are small along $\kappa_{\mathrm{p}}C_{\mathrm{p}}$ as $\delta_{\mathrm{er}}(\kappa_{\mathrm{p}}/k_{\mathrm{B}}T)= \pm 20$ and
$\delta_{\mathrm{er}}(C_{\mathrm{p}}(\mathrm{nm}^{-1})) = - 0.0003(\kappa_{\mathrm{p}}/k_{\mathrm{B}}T-82) \pm 0.001$.
For the N-BAR domains, the two density--curvature curves are not completely fitted simultaneously, even when the anisotropic model is used.
This  deviation is likely caused by a small cluster formation, and
$30\lesssim\kappa_{\mathrm{p}}/k_{\mathrm{B}}T \lesssim 60$ 
and $0.06 \lesssim C_{\mathrm{p}}(\mathrm{nm}^{-1})\lesssim 0.09$ are estimated.
If the definition of the protein area is modified,
the bending rigidity is changed as $\kappa_{\mathrm{p}}'=\kappa_{\mathrm{p}}a_{\mathrm {p}}/a_{\mathrm {p}}'$.

The experimental data were well-fitted without the side bending rigidity.
Including them, the fitness was improved for some of the conditions
but the others were not changed significantly.
Since positive and negative side curvatures can promote and suppress the tubulation, respectively~\cite{nogu16},
the estimation of the side rigidity and side curvature is important.
Recent experiments~\cite{zeno19,stac12,busc15,snea19} 
revealed that the intrinsically disordered domains of curvature-inducing proteins
play a significant role in membrane remodeling.
The disordered domains can be modeled by excluded-volume chains.
At a low protein density,
the membrane--chain interaction slightly increases the bending rigidity and spontaneous curvature isotropically 
(i.e., in both the axial and side directions of proteins)~\cite{hier96,bick01,auth03,wu13}.
At a high density, the inter-chain interactions have strong effects in protein clusters~\cite{hier96,mars03,evan03a,nogu22b}
and also between the clusters~\cite{wu13,nogu22b}.
These effects should be further examined  by the comparison of tether-vesicle experiments. 

In this study, we fitted only the density--curvature curves.
The estimation quality can be further improved
by additional fitting for other quantities.
For this purpose,
we have proposed two parameters, the axial force and the orientational degree.
They exhibit different behaviors from the density--curvature curve;
thus, comparison with the experimental results
can facilitate the determination of the mechanical properties. 

\begin{acknowledgments}
This work was supported by JSPS KAKENHI Grant Number JP21K03481. 
\end{acknowledgments}


\begin{thebibliography}{59}
\bibitem [{foo()}]{footnote1}%
  \BibitemOpen
  \href@noop {} {}\bibinfo {note} {In Ref.~\citenum{nogu22}, the formula is
  written for $-\pi/2< \theta_{\mathrm {ps}} \le\pi/2$, but the chemical
  potential and axial forces are inconsistently calculated.}\BibitemShut
  {Stop}%
\end{thebibliography}
\end{document}